\documentclass{sig-alternate}

\usepackage[
  pass,
]{geometry}
\usepackage{xcolor}
\usepackage{hyperref}
\hypersetup{
  colorlinks   = true, 
  urlcolor     = black, 
  linkcolor    = black, 
  citecolor   = black 
}
\usepackage{url}
\begin{document}

\setcopyright{waclicense}

%
%
\conferenceinfo{Web Audio Conference WAC-2018,}{September 19--21, 2018, Berlin, Germany.}
\CopyrightYear{2018} 

\title{Multi Web Audio Sequencer: Collaborative Music Making}
%
%
%
%
%

\numberofauthors{2} 
%
\author{
%
%
\alignauthor
       \name{Xavier Favory}
       \affaddr{Music Technology Group \\ Universitat Pompeu Fabra}
       \email{xavier.favory@upf.edu}
\alignauthor 
       \name{Xavier Serra}
       \affaddr{Music Technology Group \\ Universitat Pompeu Fabra}
       \email{xavier.serra@upf.edu}
}

\maketitle
\begin{sloppypar}
\begin{abstract}

Recent advancements in web-based audio systems have enabled sufficiently accurate timing control and real-time sound processing capabilities.
Numerous specialized music tools, as well as digital audio workstations, are now accessible from browsers.
Features such as the large accessibility of data and real-time communication between clients make the web attractive for collaborative data manipulation.
However, this innovative field has yet to produce effective tools for multiple-user coordination on specialized music creation tasks.
The Multi Web Audio Sequencer is a prototype of an application for segment-based sequencing of Freesound sound clips, with an emphasis on seamless remote collaboration. 
In this work we consider a fixed-grid step sequencer as a probe for understanding
the necessary features of crowd-shared music creation sessions. 
This manuscript describes the sequencer and the functionalities and types of interactions required for
effective and attractive collaboration of remote people during creative music creation activities.

\end{abstract}

%
%
%
%
%
%
%
%

\section{Introduction}
The advancement in web technologies now allows the creation of web-browser based music applications and is now able to schedule audio sources with acceptable timing accuracy. 
The Web Audio API is increasingly incorporating functionalities of music software, such as sound synthesis, mixing or processing. 
This brings new opportunities for creating music environments through the combination of web technologies, by simply using JavaScript code.
The creation of web-based music applications enables researchers and developers to promptly introduce and test novel research advances.

Nowadays, the creation and generation of audio content is widely facilitated, requiring limited resources and equipment.
Moreover, the Internet has made communication faster, what simplified the data collection process.
The growing amount of content available in online sharing platforms makes the web the ideal place for accessing and interacting with numerous varieties of sounds.
One of the most popular collaborative sound collections is Freesound\footnote{\url{https://freesound.org/}}. Freesound is an open audio collection that hosts and curates a repository of Creative Commons licensed audio samples~\cite{font2013freesound}. 
Moreover, it includes an API which allows the construction of applications using its services, 
which has resulted in people combining it with all sorts of web technologies\footnote{\url{https://labs.freesound.org/}}.
In addition, the API supports the retrieval of features extracted automatically from audio files, as well as advanced queries combining content analysis and metadata.

Finally, it is now possible to enable real-time synchronized bilateral exchange between the clients and the server, thanks to rapid communication protocols -- depending on the server bandwidth, and on the reliability of the Internet connection.
For instance, the WebSocket protocol provides a two-way communication channel that creates a communication `tube', that remains open between the client and the server~\cite{fette2011websocket}.
The browser and the server stay connected and can exchange real-time messages in both directions.
This facilitates the development of chat applications, as well as multiplayer games, in which user actions are passed to the server, and immediately transmitted to other users.

These aspects make the Web a privileged place for experimenting with interfaces that allow concurrent exploration and manipulation of shared content.
Collaborative work environments such as Google Docs, demonstrate how a web application can be exceptionally employed as a collaborative platform.
In this sense, the Web Audio API fosters the generation and processing of audio data, directly in the browser.
Finally, the Web hosts a significant amount of freely-available audio content.
%
Real-time communication technologies bring us a step forward by enabling collaborative access and manipulation of content. This fosters the development of rich shared resources, such as Freesound, contributing to the advent of the modern sharing culture~\cite{lawrence2008remix}.

This paper describes the development of a prototype for a collaborative web-based audio sequencer, which uses sounds from the Freesound database.
We first provide an overview of related works addressing the construction of collaborative web-based audio workstations.
Then, we present our implementation of the Multi Web Audio Sequencer and give some insight about what features can be appreciated by users collaborating remotely on a music creation session.
Subsequently, we explore the insights gathered from the use of our prototype, that shed some light about the features needed for a smooth and enjoyable collaborative music composition tool.


\section{Related work}
In recent years, the web has become an extensively used virtual space where people share data, interact and collaborate with other people.
These interactions include messaging other users, commenting user-generated content, or even working together collaboratively like in Google Docs.
The recent development of the Web Audio API allows users to compose music and produce sounds on the browser.
However, browser based applications cannot compete against professional quality computer music platforms which provide a number of sophisticated tools and techniques. They are often based on computationally complex signal processing algorithms which can take advantage of the full computational power offered by a machine using efficient compiled programming languages (e.g. C++).
Despite the clear limitations that browser based music platforms have, it is still a very attractive alternative for music creators. 
There is a huge community of developers dedicated to building new tools which can be accessed from any connected device.
For instance, Soundtrap\footnote{\url{https://www.soundtrap.com/}} is close to a regular Digital Audio Workstation. It gives access to common virtual instruments, sequencing tools and audio effects, directly from a browser.
Audiotool\footnote{\url{https://www.audiotool.com/}} was built in a modular way, where typical audio units can be connected together like in real life hardware-based scenarios.
Web based music applications are also reviving old emblematic electronic instruments, such as the HTML5 Drum machine\footnote{\url{https://www.html5drummachine.com/}} which is inspired by the TR-808 drum machine.

%
%
Many platforms on the web allow sharing of music content. 
Popular streaming platforms allow a music producer to directly share his productions and get feedback from people.
More recently, people have started sharing pieces of their compositions, by providing separated tracks which allow remixes and facilitate reuse of their content.
Creating music on a web browser makes the sharing of whole composition pieces easier.
For instance, the Chrome Music Lab, which aims at making learning music more accessible and fun, has recently introduced the Song Maker\footnote{\url{https://musiclab.chromeexperiments.com/Song-Maker/}} a simple but intuitive melody and pattern composition interface based on the quantized grid, similar to what some step sequencers propose. 
After creating a sequence, the user can share it by sending a URL which gives access to the composition to another user, who can reuse and transform it.
%
%

Music was originally performed in a group, combining different instrument sections, what enabled creating complex and enjoyable music pieces.
The work of Blaine and Fels is of particular interest for our work, since they studied which was the proper design for the creation of collaborative interfaces for creating music~\cite{blaine2003contexts}.
They argue that a balance between ease-of-learning, type of control (i.e. discrete versus continuous control), level of cross-modal interaction and the support for virtuosity varies for every instrument and interface.
Schnell and Matuszewski proposed different smooth systems enabling participative concerts~\cite{schnell2017playing}.
For instance, in Soundworks, the server and clients need to be synchronized and aware of what other agents are playing~\cite{robaszkiewicz2015soundworks}.
Another thing that distributed collaborative systems allow is to work on the same piece at the same time, in a more compositional approach.
Gurevich proposed an interactive music environment which allows real-time jamming by multiple users through a local network~\cite{gurevich2006jamspace}. 
In his work, he proposed a spatial metaphor characterizing the different layers of interactivity, by defining private, personal, shared and public user spaces.

In collaborative groupware systems, the action of one user must be propagated quickly to other users.
There are many challenges in designing and developing such systems: access control, social protocols, coordination of group operations, etc. \cite{ellis1989concurrency}.
According to Sun et al.~\cite{sun1998achieving}, maintaining consistency is one of the biggest challenges for the realization of real-time cooperative editing systems. 
%





\section{Multi Web Audio sequencer}
The main question that this work tries to address is which specific features should be used to allow multiple users to collaborate on a music creation task. We propose a step sequencer which allows to sequence sounds in a fixed-time grid.
Since most of the actions a user can perform are binary (i.e. switch of states), the problems related to consistency are easily managed.
The sequencer is available at \url{https://labs.freesound.org/sequencer/},
and the code is available at \url{https://github.com/Multi-Web-Audio/multi-web-audio-sequencer}.


\begin{figure}
  \includegraphics[width=\linewidth]{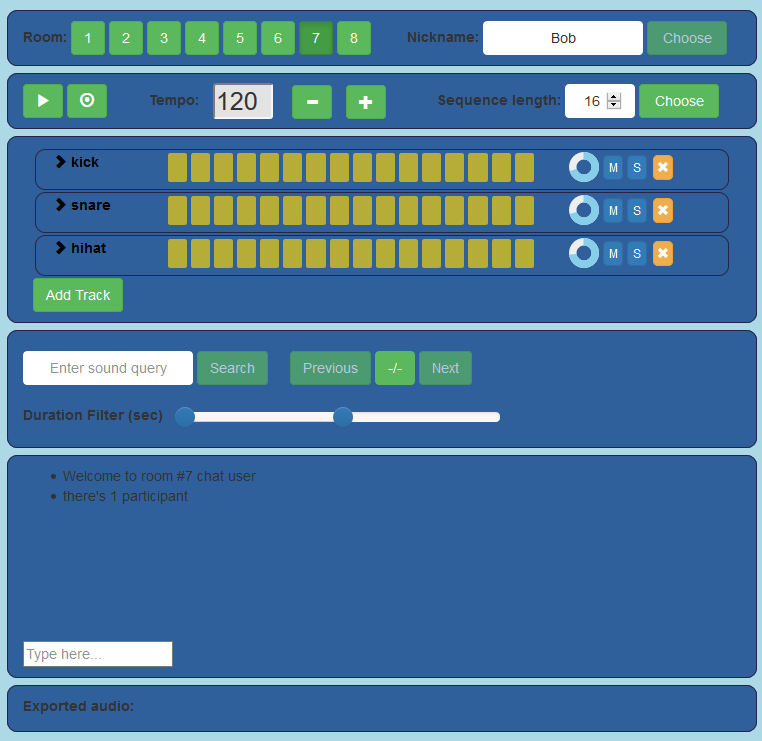}
  \caption{Initial room of the Multi Web Audio Sequencer.}
\end{figure}

\subsection{Implemented Features}
The Multi Web Audio Sequencer provides several additional features besides the basic step sequencing which allows the collaborative creation of audio loops using content from Freesound.
The sequencer runs as a client side application. All actions of a user on his browser are sent to the server which send them to other connected clients. 
Most of the available actions are shared among the users.
We present here the components of the application.

\subsubsection*{Sequencer}
The core of the application is its step sequencer. 
It allows the scheduling of samples on rounded steps of equal time-interval. Users have access to toggle buttons organized in a matrix. 
As seen in Figure 1, each line represents a different track which has an audio sample assigned.
Each column represents the time steps for which a user can activate the playing of a sample corresponding the the track line.
The figure 2 displays a collapsed track loaded with a kick audio sample.
A gain knob is accessible at the right of the pads, as well as solo, mute and delete buttons.

\begin{figure}
  \includegraphics[width=\linewidth]{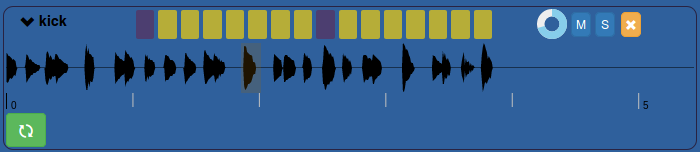}
  \caption{Collapsed track displaying the loaded audio sample waveform. The position of the shaded region correspond to sequenced audio segment.}
\end{figure}

\subsubsection*{Architecture}
A Node.js server hosts the main application. The state of the application is stored in memory on the server. Some thoughts were given about using a database to store the state, but would have led to an increased access time. However, this would be become necessary in case of a larger number of users.

\subsubsection*{Room}
Our application gives access to several rooms where any user can connect to from the home page of the application, shown in Figure 3.
Each room corresponds to a different sequencer with its own state.
When a client connects to a room, the server sends the corresponding sequencer state which is rebuilt locally.
Each of the actions performed by any user is broadcasted to all clients within the same room and update the sequencer state. We therefore had to develop our own protocol to transmit the states.
The JSON serialization format is suited to the encapsulation of data.
The major concern is to maintain consistency of the sequencer state across all the clients. Most of the actions done locally by a user are performed and sent to the server which will broadcast the actions to all the other users connected to the same room. 
The order of execution of these actions does not affect the final sequencer state.
However, for the creation and deletion of tracks actions, it is crucial to conserve the same order of execution across all the clients, since tracks are identified by their index in the session -- pushed and spliced from arrays containing tracks informations.
This is done by taking advantage of the WebSocket TCP-based protocol, which guarantee the order of delivery of the messages. Creating and removing a track takes effect only after the server receives and sends back the message to all clients, including the author of the action.

\begin{figure}[h]
  \includegraphics[width=\linewidth]{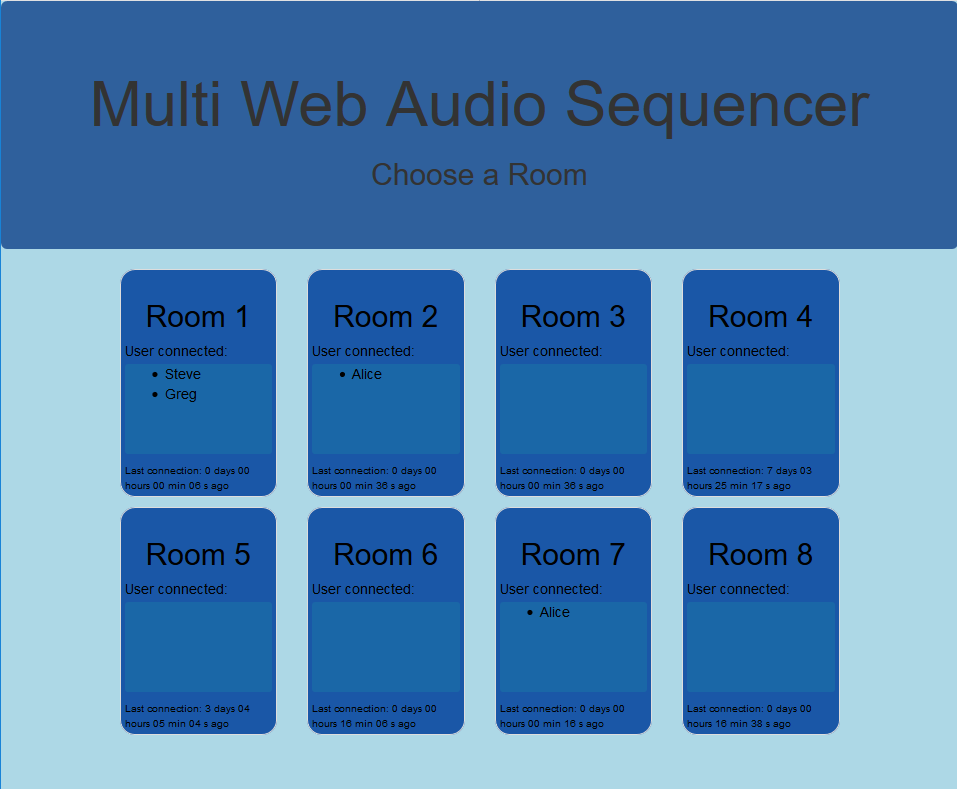}
  \caption{Home page of the application, where users can choose a room to connect to.}
\end{figure}

\subsubsection*{Chat}
A chat allows people to communicate with users connected to the same room and enables them to exchange feedback.
We also log the number of users in the room, and the individual connections.

\begin{figure}
  \includegraphics[width=0.7\columnwidth]{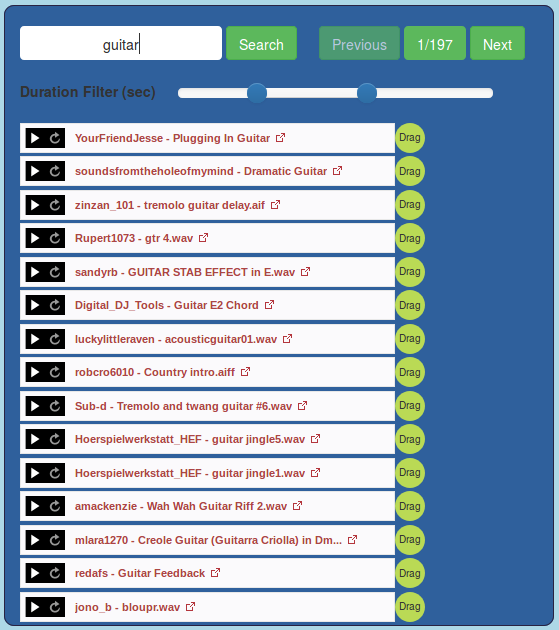}
  \centering
  \caption{Result of the query ``guitar''.}
\end{figure}

\subsubsection*{Search engine}
The text-based search engine allows people to retrieve sounds from the Freesound database. It is built using a JavaScript client for the Freesound API\footnote{\url{https://github.com/g-roma/freesound.js}}.
Metadata such as sample duration can be used to filter results, in order to find different kinds of content such as loops or single-shot samples.
Figure 4 displays the results of the query ``guitar'', which can be dragged and dropped into the sequencer tracks.





\subsection{Discussion}



Extensive evaluation of the developed prototype application has not yet been performed.
However, here we present the different aspects involved when using the proposed step-sequencer tool, i.e. when remote users connect to a Multi Web Audio room. 
The application was deployed and made available as a prototype early during its creation.
This allowed us to observe user behaviors in using shared step sequencer sessions, and therefore, we were able to find what are the must-have features for this collaborative context.
We used a first prototype of the Web Audio Sequencer as a technology probe for observing its uses in a real-world scenario, evaluating it and inspiring new ideas on novel tools for enabling its efficient use~\cite{hutchinson2003technology}. 
Our aim is to highlight the different challenges, important aspects and features to enable people who might not know each other, to start a creative music creation process concurrently.
We also observed some trends related to the use of Freesound content in a step sequencer.


\subsubsection*{Exploration of a sequencer room}
One of the major problems was that users were often lost when first arriving in a room.
The sequencer can have a lot of tracks and many curated rhythm patterns composed by different people.
New users need a way to quickly explore an ongoing session without perturbing other users in their creation process.
To alleviate this problem, we introduced the solo and mute functionalities which only take effect locally.

Users are using different types of content from Freesound. 
Distinct single-shot percussive sounds or melodic instrument notes are loaded into different tracks, which allows users to create their own loops.
The users also use loops containing repeated sections of different sound materials which they synchronize with the tempo of the sequencer.



\subsubsection*{Deadlock considerations}
At the moment, every action is broadcasted to every user. This raise some consideration about the possibility of deadlocks between users trying to modify the same element in the sequencer. Using mutexes or semaphores on the server side to protect the shared resources was considered.
Also, adding some restrictions on a track to enable the modification of it only by its owner, similar to the private and public user spaces proposed by Gurevich in~\cite{gurevich2006jamspace}, could benefit to users.

\subsubsection*{Communication between users}
We quickly understood that giving a way for users connected to a room to communicate and be aware of who was present was a mandatory feature.
Our chat and home page section list the connected members.
Moreover, the chat allows users to communicate between themselves.
Users tell each other what they are planning to do, which makes us think that logging users' activities could save them some time. Last shared actions by users could be logged, and the history could be stored.
To facilitate the exploration of a session and ease the communication between users, we could display the content type of the loaded sounds, for instance using different color on the loaded tracks.
According to Blaine and Fels, communicating user intentions in their creative choices is crucial for creating a sense of community, particularly with strangers in a public setting, for understanding their individual impact on the system is critical~\cite{blaine2003contexts}.

\section{Conclusion}
In this paper we present the Multi Web Audio Sequencer. A web application accessible from web browsers that allows people to remotely collaborate on the composition and performance of music pieces.
The simplicity of the step sequencer allows the easy overcoming of the issues in real-time collaborative systems.
We demonstrate that the web, its content, and its novel technologies allow to create innovative interactive interfaces for the manipulation of audio samples.
For future improvements, we are investigating uses of the acoustic analysis descriptors that Freesound provides, to improve the retrieval of different types of content that users might be interested in. 
For instance, enabling the possibility to filter results by pitch value for single notes or scale key for melodic loops.
We also want to make use of onset detections to automatically segment musical loops.


\section{Acknowledgments}
This work has received funding from the European Union's Horizon 2020 research and innovation programme under grant agreement No 688382 ``AudioCommons''.
The authors would like to thank Nicolas Derouineau and Valentin Nerin for their help in coding the application, and also the users of the platform who participated in creating awesome loops and gave valuable feedback.

%
\bibliographystyle{abbrv}
\bibliography{sigproc}  
%
%

\end{sloppypar}
\end{document}